\documentclass[
]{ceurart}

\usepackage{xcolor}
\usepackage{hyperref}
\usepackage{tabularx}
\usepackage{booktabs}
\usepackage[most]{tcolorbox}
\usepackage{listings}
\begin{document}

\copyrightyear{2026}
\copyrightclause{Copyright for this paper by its authors.
  Use permitted under Creative Commons License Attribution 4.0
  International (CC BY 4.0).}

\conference{CLEF 2026 Working Notes, 21 -- 24 September 2026, Jena, Germany}

\title{SourceMinds at CheckThat! 2026: NLI-Grounded Citation Auditing in a Multi-Agent Pipeline for Full Fact-Checking Article Generation}

\title[mode=sub]{CheckThat! Lab at CLEF 2026}





\author[1]{Farhan Sharukh Hasan}[%
orcid=0009-0008-1107-3287,
email=FarhanHasan@my.unt.edu,
]

\author[2]{Anirban Saha Anik}[%
orcid=0000-0002-7824-3702,
email=AnirbanSahaAnik@my.unt.edu,
]
\cormark[1]

\author[3]{Eric Liu}[%
orcid=0009-0006-5911-3394,
email=ericliu2@my.unt.edu,
]

\author[4]{Xiaoying Song}[%
orcid=0000-0001-9390-1155,
email=xiaoyingsong@my.unt.edu,
]

\author[4]{Mohotarema Rashid}[%
orcid=0009-0003-2683-7191,
email=MohotaremaRashid@my.unt.edu,
]

\author[1]{Lingzi Hong}[%
orcid=0000-0001-8412-8180,
email=Lingzi.Hong@unt.edu,
]

\address[1]{Department of Data Science,
University of North Texas, Denton, TX, United States}

\address[2]{Department of Computer Science and Engineering,
University of North Texas, Denton, TX, United States}

\address[3]{Texas Academy of Mathematics and Science (TAMS),
University of North Texas, Denton, TX, United States}

\address[4]{Department of Information Science,
University of North Texas, Denton, TX, United States}

\cortext[1]{Corresponding author.}

\begin{abstract}
This paper presents our system for Task 3 of the CLEF 2026 CheckThat! Lab, which focuses on generating full fact-checking articles from claims, veracity labels, and evidence documents. We propose a multi-agent pipeline that combines evidence retrieval, structured fact planning, article generation, gated self-critique, and NLI-based citation auditing. The system retrieves claim-relevant evidence using dense retrieval, reranking, and source-balanced selection, then generates a citation-supported article from a structured plan. A gated self-critique stage revises weakly grounded drafts, while the NLI citation auditor repairs missing citations and removes unsupported or redundant ones. The approach highlights the importance of combining evidence selection, structured generation, and post-generation citation validation for source-grounded fact-checking article generation.
\end{abstract}

\begin{keywords}
  Fact-checking \sep
  Retrieval-Augmented Generation \sep
  Multi-Agent Pipeline \sep
  Citation-Aware Generation \sep
  Natural Language Inference \sep
  Evidence Grounding
\end{keywords}

\maketitle

\section{Introduction}

The rapid growth of online misinformation has made automated fact-checking an important research direction. Existing systems commonly focus on subtasks such as claim detection, evidence retrieval, claim verification, and explanation generation \cite{guo2022survey,nakov2021automated,kotonya2020explainable}. While these components are useful for assisting fact-checkers, they do not fully address the final communication stage of professional fact-checking: writing a complete article that explains the claim, presents the evidence, and justifies the verdict for readers.

The CLEF 2026 CheckThat! Lab is a shared evaluation effort that organizes tasks to advance multilingual fact-checking  \cite{10.1007/978-3-032-21321-1_43, clef-checkthat:2026-lncs}. Task 3 addresses the gap above by requiring systems to generate full fact-checking articles from a claim, its veracity label, and a set of evidence documents \cite{clef-checkthat:2026:task3}. This task goes beyond verdict prediction because the output must be coherent, evidence-grounded, and supported with citations: a generated article must clarify the claim, identify relevant evidence, connect that evidence to the assigned verdict, and avoid unsupported statements. Prior work on fact-checking explanations and article generation shows that useful explanations require transparent reasoning and clear links between claims and evidence \cite{atanasova2020generating,zeng2024justilm,sahnan2025qraft}, which makes article generation an evidence-grounded writing problem rather than a direct prompting one.

To address this challenge, we propose a multi-agent pipeline for full fact-checking article generation. Rather than producing the article in a single step, our system separates the process into evidence retrieval, structured fact planning, article writing, self-critique, and citation auditing, so that each stage can focus on a specific part of the problem. The retrieval stage selects a compact, source-balanced evidence set; the planning stage organizes this evidence into a structured intermediate plan; the writing stage turns that plan into a citation-supported draft; a gated self-critique stage revises only weakly grounded drafts; and a natural language inference (NLI) citation auditor performs post-generation validation, repairing missing citations and removing unsupported or redundant ones. Because the pipeline uses no task-specific fine-tuning, it remains lightweight and easy to deploy while still producing source-aware articles.

We evaluate our system on the official CLEF 2026 Task 3 test set and analyze its behavior across the four official dimensions of entailment, citation precision, citation recall, and evidence coverage \cite{clef-checkthat:2026:task3}. The results indicate that decomposing the task into specialized stages helps select relevant evidence and produce citation-supported articles, while also highlighting evidence-to-claim alignment as the main remaining challenge for this kind of long-form, source-grounded generation.

\section{Related Work}

Automated fact-checking is commonly framed as a pipeline of claim detection, evidence retrieval, veracity prediction, and explanation generation \citep{guo2022survey,nakov2021automated}, with work on explainable fact-checking arguing that systems should expose not only a label but also the evidence and reasoning behind it \citep{kotonya2020explainable}. Surveys of generative language models note that while LLMs can produce such reasoning, they require strong grounding to avoid unsupported or misleading output \citep{vykopal2024generative}, which motivates our view of full fact-checking article generation as an evidence-grounded writing task rather than a label-only verification problem. Because downstream reasoning depends on the relevance of the selected passages, retrieval is central to this setting: \citet{thorne2018fever} formalized fact verification as retrieving evidence and classifying claims as supported, refuted, or not enough information, and \citet{lewis2020retrieval} extended this idea through retrieval-augmented generation by conditioning generated text on external evidence. Dense retrieval with Sentence-BERT \citep{reimers2019sentencebert} enables semantic matching between claims and passages, while cross-encoder reranking \citep{nogueira2019passage} improves relevance by jointly scoring claim--evidence pairs; we follow this retrieve-then-rerank direction but add a source-balanced selection step so that the compact evidence set is not dominated by any single document. Building on these components, several studies move beyond verdict prediction toward explanation and justification: \citet{atanasova2020generating} generate fact-checking explanations, \citet{zeng2024justilm} use retrieved evidence and fact-checking articles for few-shot justification generation, and human-centered work shows that fact-checkers need explanations that are inspectable and traceable to evidence \citep{warren2025show}. Most directly related to our setting, \citet{sahnan2025qraft} frame full fact-checking article generation as an agentic process over claims, veracity labels, and evidence documents; we share this agentic framing but use a lightweight, prompt-based pipeline whose distinguishing feature is an explicit post-generation citation auditing stage.

Citation quality is especially important for long-form fact-checking, where factual statements must be traceable to reliable sources. ALCE \citep{gao2023alce} evaluates citation-supported generation, and FActScore \citep{min2023factscore} shows that long-form outputs often mix supported and unsupported claims. \citet{madaan2023selfrefine} demonstrate that models can improve their outputs through self-generated feedback and revision, and natural language inference provides a way to test whether cited evidence entails a generated statement, building on entailment models and datasets such as RoBERTa \citep{liu2019roberta} and MultiNLI \citep{williams2018multinli}. Together, these directions motivate our combination of gated self-critique and NLI-grounded citation auditing to improve citation recall, citation precision, and factual traceability.

\section{Task Description}

\subsection{Task Overview}
We participate in \textbf{Task 3: Generating Full Fact-Checking Articles} in the CLEF 2026 CheckThat! Lab \cite{clef-checkthat:2026:task3}. The task requires systems to generate full-length fact-checking articles for input claims. Unlike standard fact verification tasks that only predict a veracity label, this task expects an explanatory article that presents the claim, discusses relevant evidence, and communicates the final factual judgment in a clear and grounded manner.

Participants are provided with a claim, its corresponding veracity label, and the evidence documents used in the fact-checking process. A complete article is generally expected to cover three main components:
\begin{itemize}
    \item \textbf{Claim background} -- a brief presentation of the claim under investigation;
    \item \textbf{Evidence discussion} -- an explanation of the relevant evidence and its relationship to the claim;
    \item \textbf{Verdict statement} -- a final conclusion that aligns with the assigned veracity label.
\end{itemize}
The task therefore requires systems to combine evidence selection, article organization, natural language generation, and citation support to produce reliable long-form fact-checking explanations.

\subsection{Task Setting}
We treat Task 3 as a long-form, evidence-grounded generation problem, in which the system must transform the provided claim, verdict, and evidence into a coherent and source-supported article. Because our approach does not use task-specific supervised training, we design a multi-agent pipeline that operates directly on the provided inputs, focusing on evidence selection, article planning, self-revision, and citation verification.

\subsection{Dataset}
\label{sec:dataset}
We evaluate on the official CLEF~2026 Task~3 test set, derived from WatClaimCheck~\cite{khan-etal-2022-watclaimcheck}. Each instance provides a claim, its metadata, a veracity label, a set of premise (evidence) articles, and a ground-truth reference article for evaluation. The test set contains 1{,}158 claims with 7.8 premise articles per claim on average. Table~\ref{tab:dataset} summarizes the statistics.

\begin{table}[h]
\centering
\caption{CLEF~2026 Task~3 test set statistics.}
\label{tab:dataset}
\begin{tabular}{lc}
\toprule
\textbf{Statistic} & \textbf{Value} \\
\midrule
Number of claims                 & 1{,}158 \\
Avg.\ premise articles per claim & 7.8 \\
Avg.\ reference length (words)   & 2{,}105 \\
\bottomrule
\end{tabular}
\end{table}

\section{Methodology}

\subsection{Problem Formulation}

We formulate Task 3 as an evidence-grounded article generation problem. Each input instance contains a claim $C$, metadata $M$, a stated verdict $V$, and a set of premise articles $D=\{d_1,d_2,\ldots,d_n\}$. From these premise articles, the system selects an evidence set $E$ and generates a full fact-checking article $A$ with source-supported explanations.

Formally, the generation process can be represented as:

\[
f(C,M,V,E) \rightarrow A
\]

where $E$ denotes the selected evidence passages retrieved from $D$, and $A$ denotes the final generated fact-checking article. The function $f$ represents the complete generation pipeline, including evidence selection, planning, article writing, revision, and citation auditing. We treat the stated verdict $V$ as a strong prior rather than an unquestioned label: the planning and writing stages are explicitly instructed to report when the retrieved evidence conflicts with $V$, rather than rationalizing it. This formulation is related to retrieval-augmented generation, where external evidence is incorporated into the generation process to improve factual grounding and reduce unsupported output \cite{lewis2020retrieval}. It also follows the broader fact verification setting, where claims are checked against textual evidence before producing a final judgment or explanation \cite{thorne2018fever}.

\subsection{Pipeline Overview}

Following the formulation $f(C,M,V,E) \rightarrow A$, our system decomposes the article generation process into retrieval, planning, writing, revision, and citation-auditing stages. The premise articles $D$ are first processed to construct a compact evidence set $E$. The selected evidence is then converted into a structured plan $P$, which guides the generation of an initial article draft $A_0$. The draft is conditionally revised into $A_1$, and the resulting article is then passed through a citation auditing stage to produce the final article $A$.

The overall pipeline can be represented as:

\[
D \xrightarrow{\text{retrieval}} E,\quad
(C,M,V,E) \xrightarrow{\text{planning}} P,\quad
(P,E) \xrightarrow{\text{generation}} A_0,
\]

\[
(A_0,E) \xrightarrow{\text{self-revision}} A_1,\quad
(A_1,E) \xrightarrow{\text{citation auditing}} A.
\]

The self-revision step is gated: rather than always invoking the critique model, the system first inspects the draft $A_0$ with lightweight citation-density heuristics and only triggers revision when the draft appears weakly grounded (see Section~\ref{sec:agents}). This decomposition makes the system more controllable than direct one-step generation. The retrieval stage selects relevant evidence, the planning stage organizes the reasoning structure, the writing stage produces the article, the revision stage improves weakly supported content, and the citation auditing stage checks whether the final article remains grounded in the provided evidence.

\subsection{Agent-Based Article Generation}
\label{sec:agents}

Our system uses a sequence of specialized agents to transform the selected evidence set $E$ into a complete fact-checking article. Among these agents, the Fact Planner Agent, Article Writer Agent, and Self-Critique Agent are prompt-based. The Evidence Retriever Agent and NLI Citation Auditor Agent do not use generation prompts; they operate through retrieval, reranking, entailment scoring, and rule-based citation processing. Overall, the system uses three main prompt templates and one optional retry prompt in the planning stage.

\textbf{Evidence Retriever Agent.} The Evidence Retriever Agent constructs the selected evidence set $E$ from the premise articles $D$. Because the evidence documents may appear in different formats, it first applies a robust loading strategy that supports structured JSON, alternate (BOM-prefixed) JSON parsing, JSONL, and raw-text extraction in cascading order until usable text is recovered, so that text can be obtained even when the evidence files are not stored uniformly. The loaded text is then cleaned, split into sentences, and grouped into short chunks using a sliding window: in our default setup each chunk spans two neighboring sentences and the window advances by one sentence, so consecutive chunks overlap by one sentence, preserving local context across chunk boundaries while keeping the retrieval units concise. The agent then selects evidence in two stages. Dense retrieval first ranks candidate chunks by their embedding similarity to the claim \cite{reimers2019sentencebert}, after which a cross-encoder reranking stage scores each claim--evidence pair more precisely to surface the most directly relevant passages \cite{nogueira2019passage}. Finally, source-balanced selection limits repeated passages from the same URL, prioritizing one chunk per distinct source before allowing additional chunks from already-seen sources. 


\textbf{Fact Planner Agent.} The Fact Planner Agent converts the selected evidence set $E$ into a structured intermediate plan $P$. Its prompt asks the model to produce the claim interpretation, stated verdict, main reason, supporting points, evidence quotes, source URLs, and possible caveats. Each supporting point must be grounded in the selected evidence. If the retrieved evidence does not fully support the stated verdict, the planner records this limitation in the caveats field instead of inventing unsupported reasoning. When the generated plan cannot be parsed as valid JSON, the system retries once with a stricter retry prompt.

\begin{tcolorbox}[
    title=Fact Planner Agent Prompt,
    colback=blue!3,
    colframe=blue!60!black,
    fonttitle=\bfseries,
    breakable
]

\textbf{System message:}

You are a careful fact-check planning agent. You read evidence and produce a structured plan that another agent will use to write an article. You output ONLY a JSON object---no preamble, no markdown fences, no commentary. Every supporting point you make must be directly traceable to text in the provided evidence. If the evidence does not support the stated verdict, say so in the \texttt{caveats} field rather than inventing support.

\vspace{0.2cm}
\textbf{User message template:}

Read the claim and evidence. Produce a fact-check plan as a single JSON object.

\textbf{Claim:} \texttt{\{claim\}}

\textbf{Claimant:} \texttt{\{claimant\}}

\textbf{Stated verdict:} \texttt{\{verdict\}}

\textbf{Evidence:} \texttt{\{evidence\_block\}}

Output this exact JSON shape: claim interpretation, verdict, main reason, supporting points, evidence quotes, URLs, and caveats.

\textbf{Rules:} Produce 3 to 6 supporting points. Each evidence quote must appear verbatim in the evidence text. Each URL must be shown in the evidence block. Never invent URLs. If the evidence is insufficient to support the stated verdict, say so explicitly in caveats. Output JSON only.

\vspace{0.2cm}
\textbf{Retry instruction:}

IMPORTANT: Your previous response could not be parsed as JSON. Output ONLY the JSON object below. Do not wrap it in markdown fences. Do not add any text before or after the JSON.

\end{tcolorbox}

\textbf{Article Writer Agent.} The Article Writer Agent generates the first article draft $A_0$ using the structured plan and the evidence index. Its prompt asks the agent to produce a 250--450 word fact-checking article in a journalistic style. The article introduces the claim, states the verdict, explains the relevant evidence, addresses caveats when necessary, and ends with a summary statement. To improve source grounding, factual sentences are required to include inline citations in the form \texttt{(source: URL)} \cite{gao2023alce}. The writer is also instructed to use only URLs from the provided evidence index.

\begin{tcolorbox}[
    title=Article Writer Agent Prompt,
    colback=green!3,
    colframe=green!50!black,
    fonttitle=\bfseries,
    breakable
]

\textbf{System message:}

You are a professional fact-checking journalist. You write 250--450 word fact-check articles grounded entirely in the provided plan and evidence index. Every factual sentence must include an inline citation in the exact form \texttt{(source: URL)}. For multiple distinct sources use \texttt{(source: URL1; URL2)}, but only when each URL contributes a different fact. Never invent facts. Never invent URLs. Use only URLs listed in the evidence index. If the plan's caveats note that evidence conflicts with the stated verdict, write what the evidence actually shows and explain the conflict; do not rationalize the stated verdict.

\vspace{0.2cm}
\textbf{User message template:}

Write a 250--450 word fact-checking article using the plan and evidence index below.

\textbf{Claim:} \texttt{\{claim\}}

\textbf{Claimant:} \texttt{\{claimant\}}

\textbf{Claim date:} \texttt{\{claim\_date\}}

\textbf{Review date:} \texttt{\{review\_date\}}

\textbf{Stated verdict:} \texttt{\{verdict\}}

\textbf{Reviewer:} \texttt{\{reviewer\}}

\textbf{Plan:} \texttt{\{structured\_plan\}}

\textbf{Evidence index:} \texttt{\{evidence\_url\_index\}}

\textbf{Structure:} Open with one sentence stating what the claim says. State the verdict clearly in the second or third sentence. Walk through the evidence supporting the verdict, one supporting point per one to two sentences. If the plan's caveats flag a conflict between evidence and verdict, address it directly. End with a one-sentence summary of why the claim is rated as it is.

\textbf{Citation rules:} Every factual sentence ends with at least one \texttt{(source: URL)} citation. A URL may be cited only if its evidence summary directly supports the sentence. Combine sources with semicolons only when each URL adds a distinct fact. Do not cite the same URL twice in one sentence. Do not output markdown headings, bullet points, or lists.

\texttt{\{exemplar\_article\}}

Now write the article for the current claim.

\end{tcolorbox}

\textbf{Self-Critique Agent.} The Self-Critique Agent revises the initial draft $A_0$ into a cleaner version $A_1$. To avoid unnecessary generation cost, the critique step is gated by a heuristic check on $A_0$: revision is triggered only when the draft shows signs of weak grounding. Specifically, when it contains fewer than three citations, draws on fewer than two distinct source URLs, or includes a substantial factual sentence with no citation at all. Drafts that pass these checks bypass revision, so that $A_1 = A_0$. When revision is triggered, the agent's prompt asks it to review the article sentence by sentence and check whether the cited evidence supports each generated statement. If a sentence is unsupported, weakly supported, or connected to an incorrect citation, the agent is instructed to revise the sentence, replace the citation, or remove the unsupported content. Revisions that collapse the article (too short, too long, or stripped of all citations) are discarded in favor of the original draft. This stage follows the idea of self-refinement, where a model improves its own output through feedback and revision \cite{madaan2023selfrefine}.

\begin{tcolorbox}[
    title=Self-Critique Agent Prompt,
    colback=orange!4,
    colframe=orange!75!black,
    fonttitle=\bfseries,
    breakable
]

\textbf{System message:}

You are a strict fact-check editor. Your job is to revise a draft article so that every cited sentence is genuinely supported by the cited URL's evidence. Output ONLY the revised article---no commentary, no preamble, no analysis section. Keep the same overall structure and length as the draft.

\vspace{0.2cm}
\textbf{User message template:}

Below is a draft fact-checking article and the evidence that was used to write it.

For each sentence in the draft, check whether the cited URL's evidence summary actually supports the claim made in the sentence. If a sentence is unsupported, weakly supported, or cites a URL whose summary does not match the sentence's claim, either rewrite the sentence so it accurately reflects what the evidence shows, replace its citation with one that genuinely supports the rewritten sentence, or remove the sentence if no available evidence supports any version of it. Keep all well-supported sentences as they are.

\textbf{Constraints:} Length must remain 250--450 words. Use only URLs from the evidence index below. Every factual sentence must end with a \texttt{(source: URL)} citation. Do not add new factual claims that go beyond the evidence. Output the revised article only---no critique notes, no diff, no JSON.

\textbf{Evidence index:} \texttt{\{evidence\_url\_index\}}

\textbf{Draft article:} \texttt{\{draft\}}

\textbf{Revised article:}

\end{tcolorbox}

\textbf{NLI Citation Auditor Agent.} The NLI Citation Auditor Agent performs post-generation citation validation. Unlike the planning, writing, and critique agents, this stage is not prompt-based. It uses natural language inference to check whether cited evidence supports the generated factual statements~\cite{williams2018multinli, reimers2019sentencebert}, and proceeds in three rule-based passes interleaved with invalid-citation filtering. First, it removes any citation whose URL does not belong to the valid premise articles. Second, it repairs missing citations: for each uncited factual sentence, it shortlists the most similar evidence chunks via embedding similarity and then verifies entailment with the NLI model, attaching a citation only when an evidence chunk entails the sentence above a confidence threshold. Third, it prunes redundant citations from sentences that cite multiple sources, removing a source when the remaining citations still entail the sentence and the dropped source adds little marginal support. Invalid-citation filtering is re-applied after each repair pass so that no out-of-scope URL survives into the final article. This staged design is intended to improve citation precision, citation recall, and factual traceability.

\subsection{Final Output Generation}

After citation auditing, the system produces the final article $A$. Each prediction is stored as a JSON object containing the claim identifier and the generated \texttt{factchecking\_article}. Overall, the methodology treats full fact-checking article generation as a controlled evidence-grounded process by decomposing $f(C,M,V,E) \rightarrow A$ into retrieval, planning, writing, revision, and citation auditing.


\section{Experiment}
\subsection{Experimental Setup}
\label{sec:experimental_setup}

\textbf{Models.} Our pipeline is inference-only and combines one generation model with three smaller supporting models. The Fact Planner, Article Writer, and Self-Critique agents share a single instruction-tuned generation model, \texttt{Qwen2.5-32B-Instruct}\footnote{https://huggingface.co/Qwen/Qwen2.5-32B-Instruct}, which we serve through a vLLM OpenAI-compatible chat endpoint rather than loading it locally. Decoupling the generator in this way frees GPU memory and lets the supporting models run on the same host. The supporting models are loaded locally and have a small footprint: dense retrieval uses the \texttt{all-mpnet-base-v2}\footnote{https://huggingface.co/sentence-transformers/all-mpnet-base-v2} sentence embedder, reranking uses the \texttt{bge-reranker-v2-m3}\footnote{https://huggingface.co/BAAI/bge-reranker-v2-m3} cross-encoder, and citation auditing uses the \texttt{roberta-large-mnli}\footnote{https://huggingface.co/FacebookAI/roberta-large-mnli} natural language inference model. The same model configuration is used for all instances; no task-specific fine-tuning is performed.

\textbf{Decoding configuration.} Because the three prompt-based agents play different roles, they use different decoding settings. The Fact Planner runs in deterministic mode (temperature $0.0$) so that the structured JSON plan is stable and parseable. The Article Writer uses temperature $0.2$ with nucleus sampling ($\text{top-}p = 0.9$) and a repetition penalty of $1.05$ to keep the prose fluent while limiting degeneration. The Self-Critique agent uses a slightly lower temperature of $0.15$ for more conservative edits. Generation length is capped at $700$ tokens for the planner, $800$ tokens for the writer, and $800$ tokens for the critique stage.

\textbf{Retrieval and selection configuration.} Evidence text is split into overlapping two-sentence chunks (sliding window, stride one). For each claim, dense retrieval shortlists up to $60$ candidate chunks, the cross-encoder reranks them, and source-balanced selection returns the final evidence set $E$ of $10$ chunks, allowing at most $2$ chunks per source URL.

\textbf{Citation auditing configuration.} The NLI Citation Auditor uses an entailment threshold of $0.5$ for both recall repair and precision pruning. During recall repair, each uncited factual sentence is matched against its top-$3$ most similar evidence chunks before NLI verification. During precision pruning, a citation is treated as redundant when the remaining citations still entail the sentence above the threshold and the dropped source changes the entailment score by no more than a margin of $0.05$.


\textbf{Infrastructure.} The generation model was served with vLLM on an H100 GPU, while the supporting embedder, reranker, and NLI models were loaded locally with a small memory footprint.

\subsection{Evaluation Metrics}
\label{sec:evaluation_metrics}
We follow the official CLEF 2026 CheckThat! Task~3 protocol \cite{clef-checkthat:2026:task3} and report the mean of four metrics. The \textbf{entailment score} is a reference-based metric computed bidirectionally between the generated article and the ground-truth reference: both are split into sentence chunks, scored with an NLI model in each direction (max over target chunks, then mean), and the two directions are averaged. \textbf{Citation recall} and \textbf{citation precision}, following Gao et al.~\cite{gao2023alce}, are computed per cited sentence: recall is $1$ if the concatenated evidence of all cited URLs entails the sentence, and precision is $1$ only if, in addition, every cited URL is necessary (removing any one breaks entailment). \textbf{Evidence coverage} is the fraction of a claim's input evidence URLs that appear as citations, independent of entailment. The final score is the mean of these four.

\section{Results and Discussion}
Table~\ref{tab:results} presents the official test set results of our submitted system alongside the shared-task baseline, using the evaluation metrics described in Section~\ref{sec:evaluation_metrics}.

\begin{table}[h]
\centering
\caption{Official test set performance of our submitted system (SourceMinds) compared with the shared-task baseline.}
\label{tab:results}
\begin{tabular}{lcc}
\toprule
\textbf{Metric} & \textbf{SourceMinds} & \textbf{Baseline} \\
\midrule
Entailment Score   & 0.245 & 0.298 \\
Citation Precision & 0.337 & 0.223 \\
Citation Recall    & 0.339 & 0.240 \\
Evidence Coverage  & 0.394 & 0.329 \\
\midrule
Mean Score         & 0.329 & 0.272 \\
\bottomrule
\end{tabular}
\end{table}

Our system achieved a mean score of $0.329$, above the shared-task baseline ($0.272$). Because the four dimensions correspond closely to the stages of our pipeline, we read the scores as feedback on individual design choices rather than as a single aggregate number, and we compare against the baseline to calibrate what each score means in absolute terms.

Our clearest gains over the baseline are in citation quality. Citation precision ($0.337$ vs.\ $0.223$) and citation recall ($0.339$ vs.\ $0.240$) both improve substantially, and their closeness reflects the two stages that jointly shape them: the article-writing prompt, which requires an inline citation on every factual sentence, and the NLI citation auditor, whose recall-repair pass attaches citations to uncited sentences while its precision-pruning pass removes redundant ones. Because the auditor's precision criterion is strict---every cited URL must be individually necessary---this stage optimizes the same quantity the metric rewards.

Evidence coverage ($0.394$) is our highest dimension and also exceeds the baseline ($0.329$), indicating that retrieve-then-rerank with per-URL selection surfaces relevant material and spreads citations across distinct sources rather than concentrating on a single document.

The entailment score is our lowest dimension ($0.245$) and the only one that falls below the baseline ($0.298$). Entailment is a reference-based measure computed bidirectionally against the ground-truth article, so it rewards reconstructing the reference's argument, not merely attaching valid citations. The gap between our strong citation scores and our weaker entailment score shows that local grounding does not imply global alignment: our articles reliably cite supporting evidence sentence by sentence, but do not fully reproduce the evidence-to-verdict reasoning of the reference. This pattern is consistent with the qualitative findings of Sahnan et al.~\cite{sahnan2025qraft}, whose expert evaluation reports that LLM-based systems tend to present factually relevant, correctly sourced information yet fail to construct the arguments that connect that evidence to the verdict.

Taken together, the results suggest that decomposing the task into specialized stages is effective for long-form fact-checking article generation: retrieval and source-balanced selection provide relevant material, while the writing and citation-auditing stages produce more traceable articles than the baseline. The entailment gap, however, indicates that tighter evidence-to-claim alignment remains the main bottleneck. Future gains should come less from whether citations are present and more from an explicit reasoning stage that organizes the selected evidence into the reference's line of argument before generation, rather than validating grounding sentence by sentence after the fact.
\section{Conclusion}

We presented a multi-agent pipeline for generating full fact-checking articles from claims, verdicts, and evidence documents. The system combines evidence retrieval, source-balanced selection, structured planning, article writing, gated self-critique, and NLI-based citation auditing, treating article generation as a controlled evidence-grounded process rather than a single-step prompting problem. Our results show that this decomposition is effective at selecting relevant evidence and producing citation-supported articles, with evidence coverage as the strongest dimension. The lower entailment score, however, shows that citations alone do not guarantee support, and that aligning generated statements with the underlying evidence is the key remaining challenge. Overall, the pipeline offers a modular and reproducible framework for source-aware fact-checking article generation.
\section{Limitations and Future Work}

Our system has several limitations that point to directions for future work. First, the same generation model is used for fact planning, article writing, and self-critique. While this keeps the pipeline simple, it may limit the independence of the revision stage, since the critic can inherit the writer's reasoning patterns; using different model families for generation and critique could make self-revision more effective. Second, the citation auditing stage relies on a single NLI model, which may struggle with long evidence passages, indirect support, or partially relevant context; stronger judge models or an ensemble of verifiers could improve reliability. Third, article quality depends heavily on the retrieved evidence, so incomplete or weakly related passages can lead to missing details or unsupported statements, motivating stronger retrieval and evidence-to-claim alignment. Finally, the pipeline uses no task-specific fine-tuning, which keeps it flexible and easy to deploy but may cap performance relative to adapted models. Future work will explore lightweight fine-tuning, preference optimization, and stronger citation validation to improve entailment, citation precision, citation recall, and evidence coverage.

\section*{Declaration on Generative AI}

During the preparation of this work, the authors used generative AI tools to assist with refining pre-written text, improving clarity and phrasing, and formatting LaTeX. After using these tools, the authors reviewed and edited the content as needed and take full responsibility for the content of the publication.

\bibliography{references}

\appendix



\end{document}